\documentstyle[12pt,aaspp4,flushrt]{article}

\def\r0{r_0}
\def\thetacr{\theta_{\rm cr}}
\def\xc{x_{\rm c}}
\def\yc{y_{\rm c}}
\def\beq{\begin{equation}}
\def\eeq{\end{equation}}
\def\thetaE{\theta_{\rm E}}
\def\Rs{R_{\rm s}}
\def\Mlens{M_{\rm lens}}

\begin{document}

\title{Gravitational Microlensing of Gamma-Ray Burst \\ 
Afterglows by Single and Binary Stars}

\author{Shude Mao$^1$ and Abraham Loeb$^2$} 

\affil{$^{1}$ Univ. of Manchester, Jodrell Bank Observatory, Macclesfield,
Cheshire SK11 9DL, UK} 

\affil{$^{2}$ Harvard-Smithsonian CfA, 60 Garden Street, Cambridge, MA
02138, USA}

\begin{abstract}
We calculate the magnification light curves due to stellar microlensing of
gamma-ray burst (GRB) afterglows. A GRB source appears on the sky as a thin
ring which expands faster than the speed of light and is maximally
magnified as it crosses the lens caustics. While a single star lens
produces a single peak in the magnification light curve, binary star lenses
may produce multiple peaks.  The shape of the magnification light curve
provides invaluable information on the surface brightness distribution of
the afterglow photosphere on sub micro-arcsecond scales.  We find that {\it
all} afterglows are likely to show variability at the level of a few
percent about a year following the explosion, due to stars which are
separated by tens of Einstein radii from their line-of-sight.

\end{abstract}

\section{Introduction}

The lensing cross-section of a star at cosmological distances is defined by
the Einstein angle,
\begin{equation}
\thetaE=\left({4GM_{\rm lens}\over c^2 D}\right)^{1/2}= 1.6 \left({M_{\rm
lens}\over 1 M_\odot}\right)^{1/2}\left({D\over 10^{28}~{\rm
cm}}\right)^{-1/2}~{\rm micro-arcsecond},
\label{eq:1}
\end{equation}
where $M_{\rm lens}$ is the lens mass, $D\equiv {D_{\rm os}D_{\rm ol}/
D_{\rm ls}}$ is the ratio of the angular-diameter distances between the
observer and the source, the observer and the lens, and the lens and the
source (Schneider, Ehlers, \& Falco 1992, p. 27).  Typical astrophysical
sources and lenses move at speeds of $v\la 10^3~{\rm km~s^{-1}}$, and hence
produce a characteristic event duration of $\sim D_{\rm os}\thetaE/v\ga 25$
years for solar mass lenses.

Coincidentally, the apparent angular size of gamma-ray burst (GRB)
afterglows is also of order a micro-arcsecond, i.e. comparable to the
Einstein angle of a solar mass lens at a cosmological distance.  Loeb \&
Perna (1998) noted that the highly relativistic expansion of these sources
would shorten considerably the duration of their microlensing events. A GRB
afterglow is predicted to appear on the sky as a narrow ring which expands
laterally at a superluminal speed, $\sim \Gamma c$, where $\Gamma\gg1$ is
the Lorentz factor of the relativistic blast wave which emits the afterglow
radiation (Waxman 1997b; Sari 1998; Panaitescu \& Meszaros 1998; Granot,
Piran, \& Sari 1999).  For $\Gamma\sim 10$ the event duration is shortened
by 3--4 orders of magnitude relative to typical astrophysical sources, and
is reduced to the convenient timescale of days.  Recently, Garnavich, Loeb,
\& Stanek (2000) have reported the possible detection of a microlensing
magnification feature in the optical-infrared light curve of GRB
000301C. The achromatic bump in the light curve is well fitted by a
microlensing event of a $0.5 M_\odot$ lens separated by an Einstein angle
from the source center, as long as the source is a ring with a narrow
fractional width ($\sim 10\%$), in accordance with earlier theoretical
predictions. In a case similar to GRB sources, Koopmans \& de Bruyn (2000)
have recently reported evidence for intra-day radio microlensing in the lens
B1600+434 due to superluminal motions of the lensed source. On a
longer-time scale, microlensing
of the lensed QSO 2237+0305 had been observed over many years and more
recently in real-time (Wozniak et al. 2000 and references therein).

The predicted angular size of GRB afterglows was already confirmed through
the detected transition from diffractive to refractive scintillations in
the radio afterglow of GRB 970508 (Goodman 1997; Waxman, Kulkarni, \& Frail
1998).  Unfortunately, scintillations are not a sufficiently delicate probe
to provide information about the ring-like surface brightness distribution
of a GRB source on the sky.  On the other hand, microlensing can resolve
the sub micro-arcsecond structure of GRB photospheres as a function of
observed photon frequency.  In this {\it Letter}, we expand the original
discussion of Loeb \& Perna (1998) on single star lenses, and calculate the
magnification light curve for two other lens configurations, namely binary
star lenses and a single star in an external shear field (due to the
gravitational potential of the host galaxy or a distant binary
companion). These lensing configurations are more likely to describe stars
which are embedded in the luminous cores of galaxies. At cosmological
distances, the simple model of single star lensing is adequate only for
stars which reside in the outer regions of galaxies or in the intergalactic
medium.

The probability for having an intervening lens star within a projected
angular separation $\theta$ from a source at a redshift $z\sim 2$ is~~$\sim
0.3 \Omega_\star (\theta/\thetaE)^2$ (Press \& Gunn 1973; Blaes \& Webster
1992; Nemiroff 1998; Nemiroff et al. 1998), where $\Omega_\star$ is the
cosmological density parameter of stars.  The value of $\Omega_\star$ is
bounded between the density of the luminous stars in galaxies and the total
baryonic density as inferred from big bang nucleosynthesis, $7\times
10^{-3} \la \Omega_\star\la 5\times 10^{-2}$ (Fukugita, Hogan, \& Peebles
1998). Hence, all GRB afterglows should show evidence for events with
$\theta \la 20\thetaE$, for which microlensing provides a small
perturbation to the light curve\footnote{This crude estimate ignores the
need to subtract those stars which are embedded in the dense central
regions of galaxies, where macrolensing dominates and the microlensing
optical depth is of the order of unity.}. In this {\it Letter}, we
calculate the magnification light curves of such events, as well as light
curves of rarer events with a smaller impact parameter -- for which the
peak amplitude is larger.

In \S 2 we describe the method of our calculation, and in \S 3 we present
our numerical results. Finally, \S 4 summarizes the main implications of
these results.

\section{Method of Calculation}

We assume that the GRB source appears on the sky as an inner disk with a
uniform surface brightness bounded by a uniform outer ring. This model has
the minimum number of free parameters, which are necessary to describe
results from detailed calculations (see Figs. 11 and 12 in Granot et
al. 1999). We specify the ratio between the surface brightness of the inner
disk and that of the outer ring using a contrast parameter, $C$. In
reality, the contrast $C$ and the ring fractional width $W$, depend on
photon frequency and obtain different values for the different spectral
slope regimes that are separated by spectral breaks in an afterglow
spectrum (Sari, Piran, \& Narayan 1998; Granot et al. 1999).  The outer
boundary of the ring is determined by the sharp cut-off in relativistic
beaming at angles $\ga 1/\Gamma$ relative to the center of the
explosion. The dimming of the inner disk of the image results from the fact
that the outer ring suffered a longer geometric time delay and hence must
have been emitted at earlier times -- when the fireball was brighter.

The ring-like image of the source expands outwards at a superluminal speed,
and so the lens is assumed to be stationary.  During the relativistic
expansion of a spherical fireball, the apparent source radius increases
with time as $\propto t^{5/8}$ (Waxman 1997b; Loeb \& Perna 1998).  The
effects of lensing on the source image depend only on its angular structure
in units of the Einstein angle, $\thetaE$; this angle, given in
equation~(\ref{eq:1}), is a function of the source and lens redshifts and
the lens mass. Hence, we normalize all angular scales in units of $\thetaE$
and parametrize the time-dependent angular radius of the ring as,
\begin{equation} \label{source}
\Rs(t) = R_0 t^{5/8}, 
\label{eq:R}
\end{equation}
where $R_0$ is the angular radius of the outer ring in units of the
$\thetaE$ after one day, and $t$ is the time in days.  For binary lenses,
we take $M_{\rm lens}$ to be the total mass of the system.  For
definitiveness, we adopt a value of $R_0=0.5$, as found by Garnavich, Loeb
\& Stanek (2000) for GRB 000301C.  The time axis in our plots can be
trivially rescaled by $R_0^{-8/5}$ for other choices of $R_0$.

We note that equation~(\ref{eq:R}) is adequate only as long as the GRB
fireball is spherically-symmetric and highly relativistic. A collimation of
the outflow into a jet changes this scaling as soon as the fireball Lorentz
factor decelerates to a value smaller than the inverse of the collimation
angle (Rhoads 1997).  Eventually, the total energy release is isotropized
(as if it came from a quasi-spherical explosion) after the fireball enters
the non-relativistic stage of its expansion. Using the geometric time-delay
of the ring and the Blandford-McKee (1976) self-similar solution for the
expansion of a relativistic spherical fireball, one finds that the Lorentz
factor of the blast wave at an observed time $t$ is given by,
$\Gamma\approx 6.4 (E_{52}/n_0)^{1/8} t^{-3/8}$, where $E_{52}$ is the
(isotropically-equivalent) energy release in units of $10^{52}~{\rm
erg~s^{-1}}$ and $n_0$ is the number density of the ambient gas in units of
$1~{\rm cm^{-3}}$ (e.g., Waxman 1997a,b). Hence, the shock wave is expected
to become non-relativistic several months after the explosion.  In the
non-relativistic regime, the source follows the Sedov-Taylor self-similar
scaling of $\Rs\propto t^{2/5}$, and the ring contrast is determined solely
by the limb-brightening effect.  Since no calculations were done so far to
describe the evolution of the source image on the sky for collimated
outflows or during the transition to the non-relativistic regime, and since
the power-law index of $2/5=0.4$ is not very different from $5/8=0.625$, we
assume for simplicity that the fireball is spherically-symmetric and adopt
equation~(\ref{eq:R}) as a crude approximation for the expansion of the
source on the sky up to a year after the GRB trigger.  The detailed
evolution of the source image requires elaborate numerical calculations,
and its magnification by a simple point mass lens will be considered
elsewhere (Granot \& Loeb 2000).  Here we adopt a simple source model and
focus on the properties of more complex lens systems.

We consider three types of lenses: a single point-mass, a single point-mass
with external gravitational shear, and a binary star lens.  The effect of
the host galaxy on the microlens is well described by the second case.  The
third case corresponds to a situation when two stars are sufficiently close
together so that they act coherently as a binary lens. About forty binary
lensing events have been detected so far in microlensing searches in the
local group (Alcock et al. 2000a; Udalski et al. 2000).

For the single star lens, we choose the origin to be at the lens position
and define the source properties by $R_0, W$, and $C$.  The angular
separation between the lens and the source center is parametrized by
$b \equiv \theta/\thetaE$. For the star$+$shear case, the lens is
located again at the origin, and the shear is oriented along the $x$-axis,
with its strength given by $\gamma$ (e.g., Mao 1991). 
For a primary lens
$\Mlens$, a perturbative lens of mass $m$ at distance $r$ will provide 
a shear $\gamma=m/r^2$, where $m$ is in units of $\Mlens$
and $r$ is in units of the Einstein radius corresponding to mass $\Mlens$.
For the binary case, the two lenses are parametrized by a mass ratio, $q$,
and separation, $a$, with the origin centered at the mid-point of the
lenses. In both cases, the source center is specified by some point $(\xc,
\yc)$.

In all three cases, the lens equation can be manipulated into a complex
polynomial using a complex coordinate notation (Witt 1990; Mao 1991; Witt
\& Mao 1994). The associated polynomial can be readily solved using
well-known numerical schemes (e.g., Press et al. 1992) to yield the image
positions and magnifications for any source position.  Given our assumed
source profile, the magnification is given by (see Eq. 8 in Loeb \& Perna
1998)
\begin{equation} \label{mu}
\mu(\Rs, W, C) = {\Psi(\Rs) -(1-C) (1-W)^2 \Psi[(1-W)\Rs] \over
1-(1-C)(1-W)^2, }
\end{equation}
where $\Psi(\Rs)$ is the magnification for a uniform surface-brightness
disk of radius $\Rs$.  While the calculation of $\Psi(\Rs)$ is
straightforward for the single star case (e.g., Witt \& Mao 1994; Schneider
et al. 1992, p. 313), it is no longer simple for the star$+$shear case and
for the binary lens case. For these two cases, one has to integrate over
the singularities in the magnification amplitude as the expanding source
sweeps across the lens caustics. Equation~(\ref{mu}) only requires
knowledge of the magnification for a uniform source of radius $R$ (where
$R=\Rs$ or $(1-W)\Rs$).  Since gravitational lensing conserves surface
brightness, the magnification of each image is simply the ratio of the
image area to the source area, $\pi R^2$.

In order to find the image area using Stokes theorem, we only need to solve
the lens equation for the mapping of the confining circle (Gould \&
Gaucherel 1997; Dominik 1998).  We therefore place $n$ points on the circle
and distribute them uniformly in angle. For each point we then solve the
lens equation to obtain the corresponding image positions.  We then connect
the image positions into continuous image tracks. In general, there are
several non-crossing, closed, image tracks.  The two insets in Fig. 3 show
two examples of image tracks for a binary lens configuration (see \S3 for
lens parameters). In each inset, the thin solid line with cusps is the
caustic.  A point source located on the caustic would be infinitely
magnified with its images projected on the critical curve, shown as the
dashed line. The left inset presents the image tracks for a source center
which is located inside the caustic. In this case, the confining circle of
the source (dotted line) is mapped into five disjoint image tracks (with
the central track being highly demagnified and hence too small to be
seen). The right inset shows a source that intercepts the caustics. In this
case, the confining circle is mapped into two tracks, one of which is
embedded within the other.  If the image tracks are not embedded within
each other (such as the case shown in the left inset) then the total
magnification is simply the ratio between the total area they confine and
$\pi R^2$. However, for image tracks that are embedded inside each other,
special care needs to be applied.  If an image track is confined within
other image tracks for an odd (even) number of times, then one needs to
subtract (add) the area confined by the track. For example, in the right
inset the total area is found by subtracting the area enclosed by the inner
track from the area enclosed by the outer track.  To achieve convergence,
we double the number of points on the circle to $2n$ until the
magnification value is within 0.01\% of that found in the previous
iteration. In this {\it Letter}, we only show the magnification
history. The observed light curves can be easily obtained by multiplying
the unlensed (power-law) afterglow light curve with the magnification
amplitude at each time.

\section{Results}

The microlensing probability scales as the square of the impact parameter
from the line-of-sight to the source.  For $\Omega_\star\sim 0.01$, most
GRB afterglows would be separated by several tens of $\thetaE$ from a
star. Figure 1 shows the magnification light curves of a single star for
$b=10$ and $b=20$.  In such cases, the magnification light curve generically
gives a peak amplitude of a few percent about a year after the GRB
trigger. By considering a large statistical sample of GRBs, one may be able
to identify a systematic feature of this type and constrain $\Omega_\star$.
However, it is important to remember that the predicted microlensing
signature may change considerably at late times for collimated outflows.

The magnification light curves depend on the surface brightness
distribution of the source image.  Figure 2 shows the light curves for
different choices of $W$ and $C$ with an impact parameter $b=1$.  The three
solid curves show the magnification for $C=0$ and the three cases of
$W=0.05, 0.1$ and 0.2.  As the ring width increases, the peak magnification
drops but the full width at half maximum increases. The dotted and dashed
lines show two other examples for $C=0.5$ and 1, respectively. In both
cases, $W=0.05$.  Compared with the case of the same $W$ but with $C=0$
(solid line with the highest peak magnification), it is evident that the
increase in $C$ from 0 to 0.5 broadens the magnification curve
considerably, although the additional increase from 0.5 to 1 results in a
much milder effect.

Figure 3 shows two examples of the more complex light curves that result
from binary star lenses. The two lenses have equal-mass and a separation
$a=0.8$, and we adopt $W=0.1$ and $C=0$ for the source.  The insets in
Figure 3 illustrate the lens positions, and the resulting caustics and
critical curves. For the left inset, the source center is at the origin,
while for the right inset the source center is at $(-0.16, -1)$. For each
inset, the image tracks for the dotted circle are indicated by thick
solid lines (see \S 2).
The light curves show multiple peaks due to different caustic
crossings. Just as found in Galactic binary microlensing events (Alcock et
al. 2000a; Udalski et al. 2000, and references therein), the set of
possible light curves is diverse.  However, the magnification light curves
of GRB afterglows obtain smaller peak amplitudes and do not exhibit
features as sharp as their Galactic counterparts. These differences result
from the different source profiles; while Galactic stars have radii ($\sim
R_\odot$) which are three orders of magnitude below the Einstein radius of
their lenses, GRB afterglows obtain a size which is comparable to the
Einstein radius and hence smooth the magnification pattern over their
extended image.

Finally, Figure 4 shows the effect of an external shear on the
magnification light curve with $W=0.1$ and $C=0$.  The solid lines show
three cases where the source center is at $(\xc,\yc)=(1,0)$, with the shear
amplitude increasing in strength from 0 to 0.2. For $\gamma \la 0.1$, the
shear changes the initial magnification mildly, while for $\gamma=0.2$, the
magnification actually drops during the initial expansion phase. Since the
shear breaks the spherical symmetry of a single lens, we show by the dotted
and dashed lines the magnification light curves for $\gamma=0.1$ and 0.2
respectively, but with the source center located on the $y$ axis at
$(\xc,\yc)=(0,1)$. Compared with the case with no shear, the light curve
changes are relatively minor; they show a slight decrease in the initial
magnification followed by an offset in the time of peak magnification.  For
a single star residing in an isothermal galactic potential of a 1D velocity
dispersion $\sigma$, the external shear is $\gamma = \thetacr/(2 \theta)$,
where $\theta$ is the angle between the line of sight and the galaxy
center, and $\thetacr=4\pi (\sigma/c)^2 D_{\rm ls}/D_{\rm os}$ is the critical
angle for multiple imaging, which is typically $\la 1$ arcsecond (see
Schneider et al. 1992, \S 8.1.4). Hence, unless the line-of-sight of the
GRB passes through the central region of an intervening galaxy, the shear
is likely to be much smaller than unity and hence its effect on the
magnification light curve would be modest.

\section{Discussion}

We have found that all GRB afterglows are likely to show a microlensing
amplification bump of a few percent after about a year (Fig. 1).  The peak
magnification and its timing depend strongly on the impact parameter.  A
smaller impact parameter results in an earlier and a higher peak
magnification. For $b\ga 2$, the peak magnification is usually reached when
the limb of the source crosses the lens. Equation (\ref{source}) implies
that the time of peak magnification is then $t_{\rm peak}\approx
(b/R_0)^{8/5}$. Numerically, we find that the peak magnification scales
roughly as $\mu_{\rm peak} \approx 1+2.5 (W/0.05)^{-1/2} b^{-3/2}$ for
$C=0$. The {\it Swift} satellite \footnote{see http://swift.sonoma.edu/},
planned for launch in 2003, will provide trigger and good localization for
hundreds of GRB afterglows per year.  Identification of the statistics of
such features can be used to constrain $\Omega_\star$.  A remaining
important question is how the optical depth for microlensing (or the
associated shear) is distributed along the line of sight.  While stars in
the halos of galaxies are likely to be in the regime of very small optical
depth, the outer disk or spheroid stars will be surrounded by a modest
optical depth with a significant shear, while stars in the cores of
galaxies are in the high optical depth regime where the GRBs are likely to
be multiply imaged.  The relative abundance of stars in these populations
can be, in principle, calibrated based on ongoing microlensing studies of
the Milky Way galaxy (Alcock et al.  2000b, and references therein).

The detailed shape of the magnification light curve depends on the
contrast, $C$, and fractional width, $W$, of the ring-like afterglow image
(Fig. 2).  The values of these parameters should change 
in a predictable way across 
spectral breaks (e.g. when comparing radio to optical-infrared data), 
because of the corresponding change in the structure of the
source image on the sky (Granot et al. 2000). Detailed multi-frequency
monitoring of a microlensing event, similar or better than the data used by
Garnavich et al. (2000), could provide invaluable information on the sub
micro-arcsecond structure of GRB photospheres.

We have also found that binary lenses produce multiple peaks in the
magnification light curves (Fig. 3). However, an external shear has only a
modest effect on the magnification history, changing its early time
amplitude from the constant value which characterizes the zero shear case,
but not affecting much its late evolution. While the zero shear lens does
not resolve the source when the angular size of the source is much smaller
than its impact parameter relative to the lens, the star$+$shear lens
provides information about the expansion of the source even at these early
times. The light curves of both the star$+$shear lens and the binary lens
are determined by a small number of parameters.  For example, the binary
light curves are fixed by two binary parameters (separation and mass ratio)
and five parameters that describe the source properties: the source center
position $(\xc, \yc)$, $R_0$, $W$ and $C$ [see Eqs. (\ref{source}) and
(\ref{mu})].  Similarly to the known binary microlensing events toward the
Galactic and Magellanic Clouds, it should be possible to infer the lens and
source parameters such as $W$, $C$ and $R_0$ from a densely-sampled light
curve. The values of these  parameters at various photon frequencies 
would in turn provide new constraints on
the energy output and ambient gas density of GRB fireballs,
and hence on the central engines that power them.

\begin{acknowledgements}

This work was supported in part by NASA grant NAG 5-7039 and NSF grant
AST-9900877 for AL and by NASA grant NAG5-7016 for SM. We thank Leon
Koopmans for helpful discussions, and are grateful to Scott Gaudi, Bohdan
Paczy\'nski and the anonymous referee for useful comments on the paper.

\end{acknowledgements}

\newpage

\begin{figure}
\centerline{\epsfysize=6.5in \epsfbox{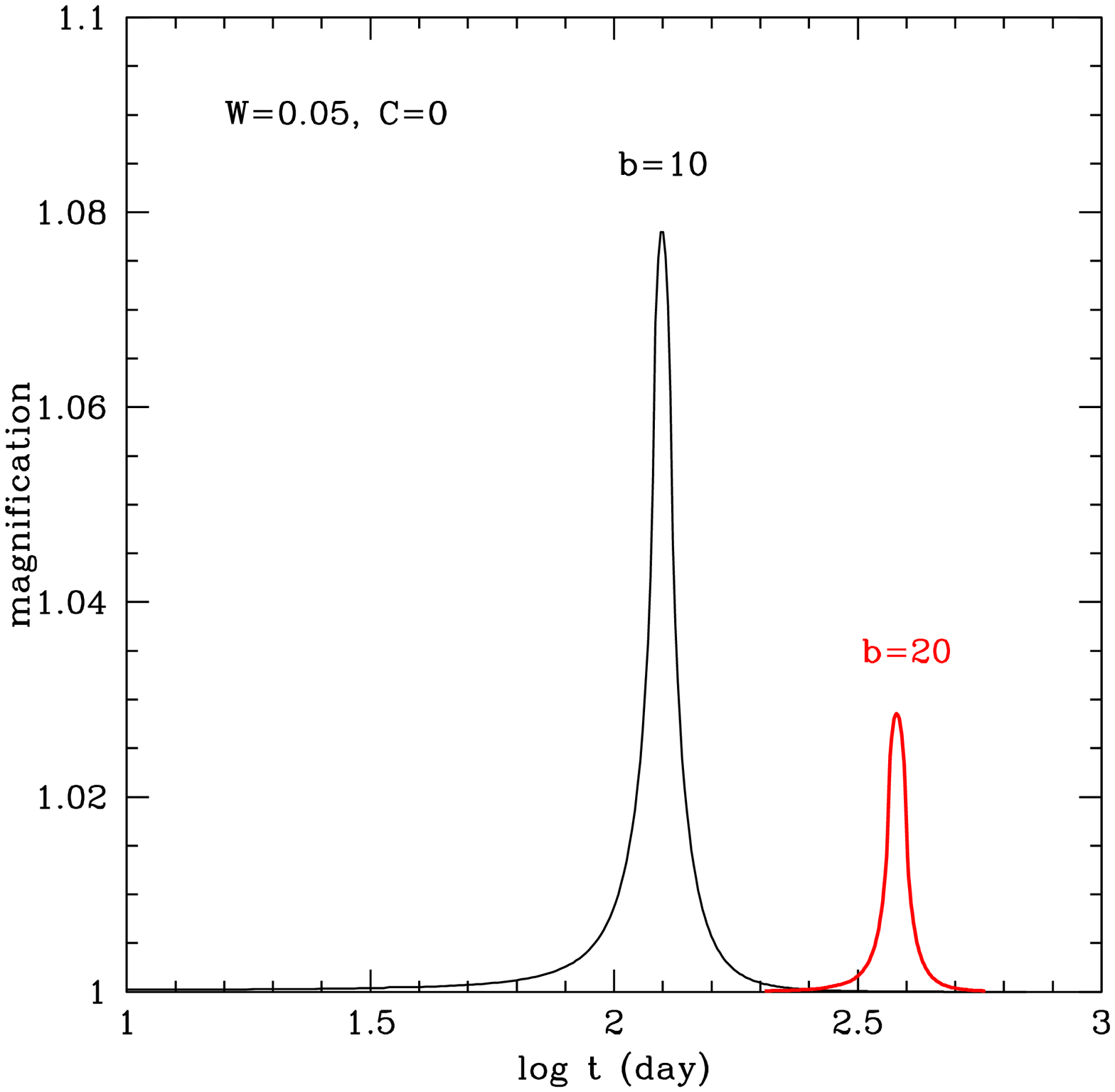}}
\caption{The afterglow magnification as a function of time for a single
star lens. The two curves correspond to an impact parameter of the source
center relative to the lens of $b=10$ and $20$, in Einstein angle
units. Such impact parameters should be typical for any GRB, given the
density of stars in the Universe.  The GRB source is assumed here to be a
narrow ring with a fractional width $W=0.05$.  }
\end{figure}

\begin{figure}
\centerline{\epsfysize=6.0in \epsfbox{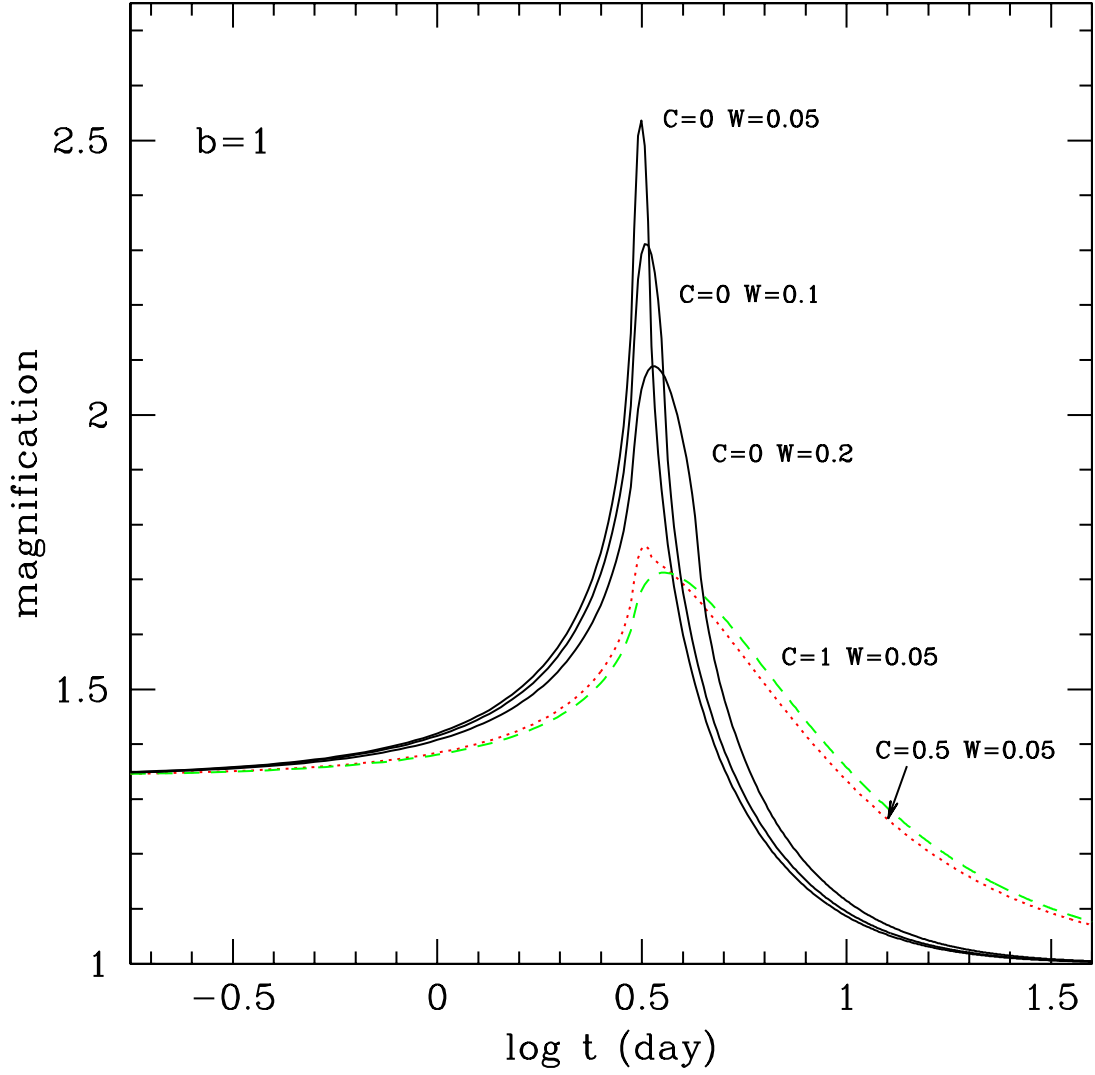}}
\caption{The magnification history for different source
profiles and a single star lens at an impact parameter $b=1$. 
}
\end{figure}

\begin{figure}
\plotone{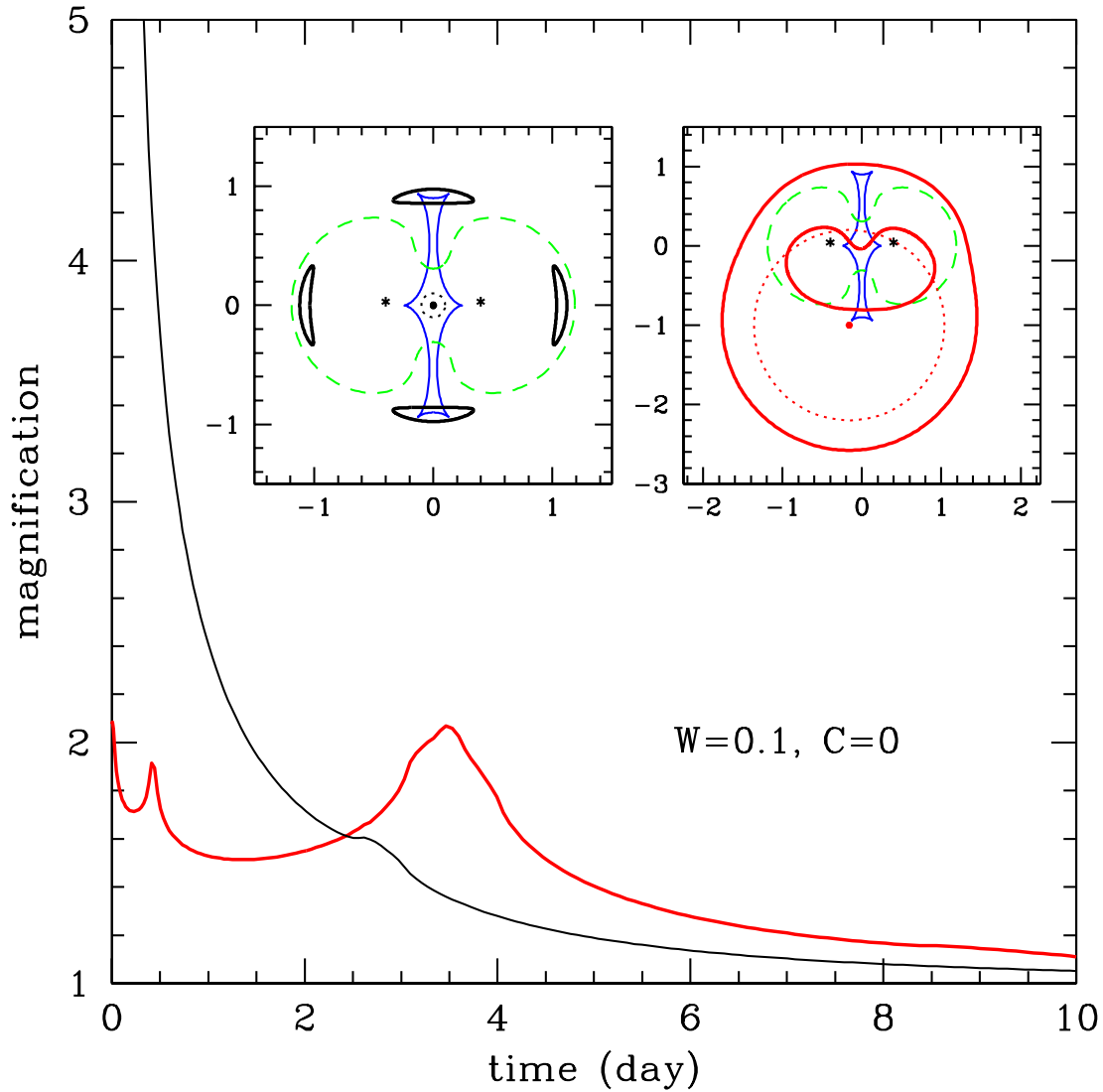}
\caption{The magnification history for a binary lens with equal mass stars
at a separation of $a=0.8$ in Einstein angle units.  The two insets show
different cases for the location of the source center (filled dots).  In
the left inset, the source center is inside the caustic region, while in
the right inset the source is close to but outside the caustic.  The
positions of the two lens stars are labelled by star symbols. The thin
solid line with cusps is the caustic while the dashed line is the critical
curve. The image tracks for the dotted circle are shown as thick solid
lines in the insets.  The thin and thick lines in the magnification light
curves correspond to the left and right insets, respectively. Note that the
location of the origin and the axis scales are different in the two
insets. }
\end{figure}

\begin{figure}
\plotone{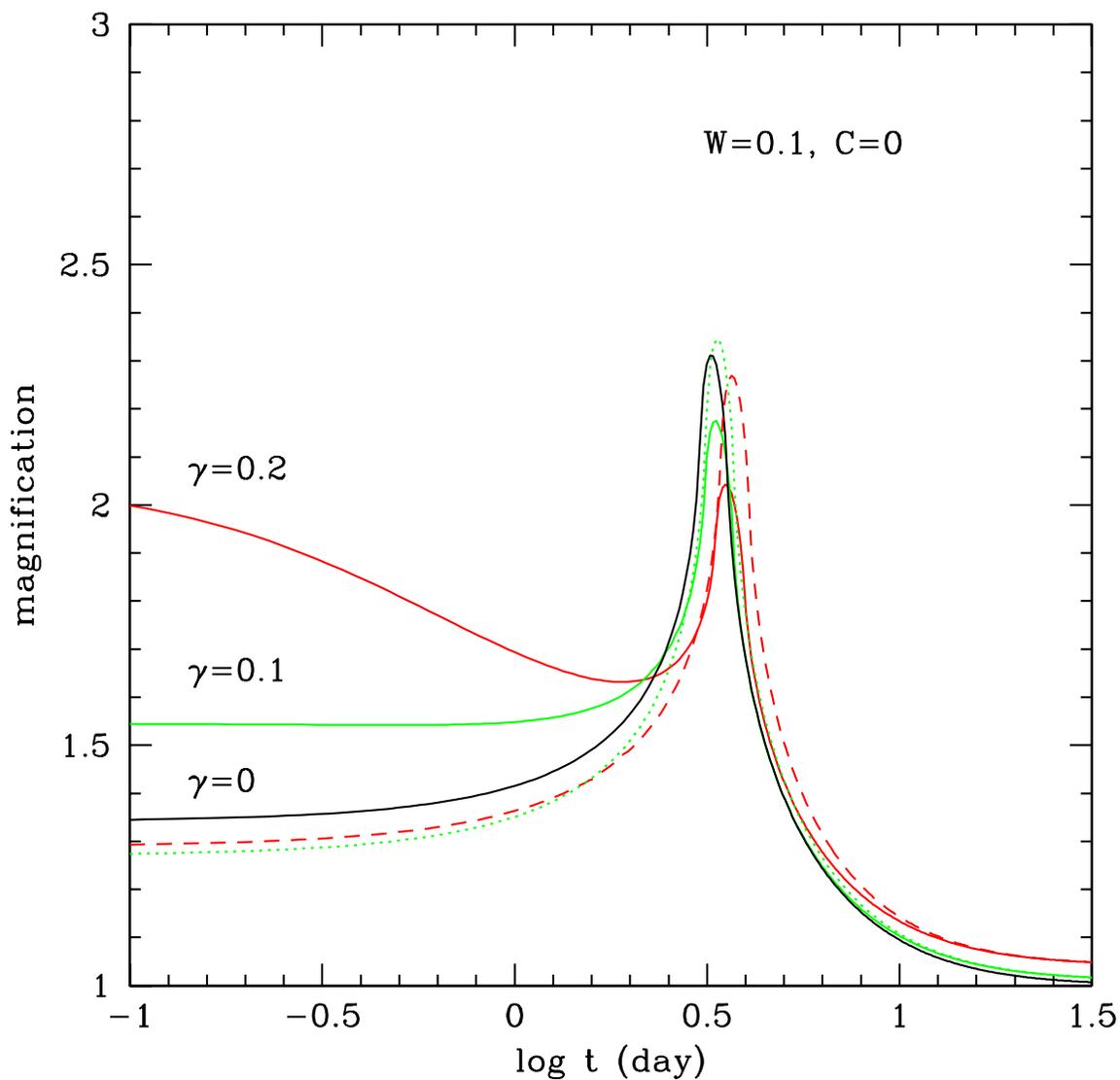}
\caption{The magnification history for a single star embedded in an
external shear, $\gamma$.  The three solid lines are for $\gamma=0, 0.1,
0.2$, respectively and with the source center on the $x$-axis, $(\xc,
\yc)=(1, 0)$. The dotted and dashed lines have $\gamma=0.1$ and 0.2,
respectively, with the source center on the $y$-axis, $(\xc, \yc)=(0, 1)$.
In all cases, we take $W=0.1, C=0$.}
\end{figure}
\end{document}